\documentclass[11pt]{article}
\DeclareUnicodeCharacter{2082}{$_2$}
\usepackage[utf8]{inputenc}
\usepackage{amssymb}
\usepackage{amsmath}
\usepackage{array}
\usepackage{graphicx,hyperref}
\usepackage[export]{adjustbox}
\usepackage{xcolor}
\usepackage{subcaption}
\usepackage{float}
\title{\textbf{Emergent Electronic Flat Bands Through Dislocation  \mbox{Defect Phase Patterning: Effective One-Dimensional Model\footnote{
To appear in Journal of Applied Mechanics (\href{https://doi.org/10.1115/1.4070026}{DOI: 10.1115/1.4070026})}
}}}

\author{
    \textbf{Aziz Fall} \\
    \small Department of Mechanical Engineering, Carnegie Mellon University \\
    \small \texttt{afall@andrew.cmu.edu} \\[1em]
    \textbf{Kaushik Dayal} \\
    \small Department of Civil and Environmental Engineering, Carnegie Mellon University \\
    \small Center for Nonlinear Analysis, Department of Mathematical Sciences, Carnegie Mellon University \\
    \small Department of Mechanical Engineering, Carnegie Mellon University
}

\begin{document}
\date{}
\maketitle

\begin{abstract}
    Recent theoretical work has predicted that dislocation patterning induces anisotropic flat bands in the electronic band diagram, which can lead to unusual effects such as unconventional superconductivity.
    This work develops a reduced-dimensional framework to provide insights into their origin. 
    An effective one-dimensional dislocation potential is constructed by averaging over the spatial distributions of dislocations along a singular direction. 
    The resulting model introduces a parameter that quantifies the strain modulation, thereby providing a transparent approach to analyze the role of dislocation strain in leading to flat band formation. 
\end{abstract}

\section{Introduction}

Flat bands are electronic dispersion relations that are independent of momentum, giving rise to an almost infinite electronic effective mass. When such bands occur near the Fermi energy, they produce strongly correlated phenomena and exotic quantum phases such as unconventional superconductivity, Mott insulators, and non-Fermi liquid behavior \cite{karmakar2025correlated,Hu2023CorrelatedFlatBands,Gao2023SingleBandMott,Huang2024NonFermiPyrochlore,Kumar2021FlatBandNFL,Sayyad2020FlatBandPairing}. Flat bands have also been associated with superfluid transport, fractional quantum Hall effects, and spin liquids \cite{tovmasyan2016,hu2023,chan2022,roy2019,herzog-arbeitman2022,wang2012}. These correlations make flat band physics an important area of study.  

Strain engineering, particularly non-uniform strain, has played a major role in the manipulation and induction of flat bands. For example, in-plane biaxial strain in bilayer graphene has been shown to produce flat bands at non-magic angles \cite{Li2025}, while uniaxially periodic strain can generate flat bands in graphene \cite{andrade2023,meng2024,GarciaMunoz2025GrapheneStrain,Wan2023FlatChernGraphene}. Heterostrain has also been demonstrated as a means of inducing flat bands in transition metal dichalcogenides (TMD) \cite{bi2019}.  

This study focuses on the role of dislocation strain fields in the formation of flat bands. 
Dislocations are a common type of line defect in crystalline materials and play a major role in functional properties such as superconductivity, thermal transport, magnetism, and optical behavior \cite{li2019quantized,fogel2001,li2017tailoring,podolyak2019,breio2024,azhar2022}.
In particular, the chemical environment around a dislocation is significantly different from the bulk perfect crystal, and can be considered as a distinct defect phase \cite{korte2022defect}.
Patterned arrays of dislocations can further lead to unusual collective behavior such as the formation of flat bands \cite{Fall2025_FlatBands}.
Because dislocations are so prevalent and influential, understanding their role in flat band physics may enable the engineering of defects to design new solid-state devices with specialized functionalities.

In our previous work \cite{Fall2025_FlatBands}, we showed that dislocation strain can induce anisotropic flat bands in the renormalized dispersion relation of a free electron gas. 
Here, we extend that study by reducing the dimensionality of the system to one dimension and developing an effective one-dimensional dislocation potential. Specifically, we derive this potential by averaging over spatial distributions of dislocations along a singular direction. The resulting model introduces a parameter that quantifies the strength of the strain modulation, providing a simplified but flexible framework for analyzing how dislocation strain drives flat band formation. This simplified description complements our earlier results and offers a clear foundation for future studies of dislocation-engineered electronic states.

\section{Motivation for an Effective Model}

We aim to better understand the origin of the flat bands observed in our previous study on dislocation-induced flat band formation under specific strain fields and directions, conditions we referred to as ``magic parameters" \cite{Fall2025_FlatBands}. Inspired by recent work in graphene \cite{andrade2023}, which showed that flat bands can form when the system is effectively reduced to one dimension under an oscillating strain with a wavelength slightly mismatched from the sublattice spacing, we hypothesize a similar mechanism in our model. Specifically, we propose that flat bands arise when the average spacing between strain modulations, i.e., the distance between high and low strain regions, approximately matches the average nearest-neighbor distance between dislocation cores with the same Burgers vector on different dipoles along a given direction. To investigate this hypothesis, we map our system onto an effective one-dimensional model and examine how different dislocation strain profiles influence the band structure, particularly when the dislocations are arranged to mimic certain features of a honeycomb lattice.

To make our discussion more precise, we begin by defining key terms relevant to the system under consideration. We consider a system composed of an array of dislocation dipoles, where all dislocation lines are aligned along the \( z \)-axis and each dipole has a fixed separation along the \( y \)-axis. In this setup, all dislocations have Burgers vectors of the same magnitude, and only two possible directions are allowed: \( \mathbf{b} \) and \( -\mathbf{b} \).

We define a \textbf{dislocation core type} based on the direction of the Burgers vector. A core with Burgers vector \( \mathbf{b} \) is classified as a \textbf{type A} dislocation core, while a core with \( -\mathbf{b} \) is classified as a \textbf{type B} dislocation core. A \textbf{dislocation dipole} consists of one core of each type, separated by a fixed vector \( \mathbf{R}_{\text{dip}} \).

\begin{figure}[H]
    \centering
    \includegraphics[width=0.35\textwidth]{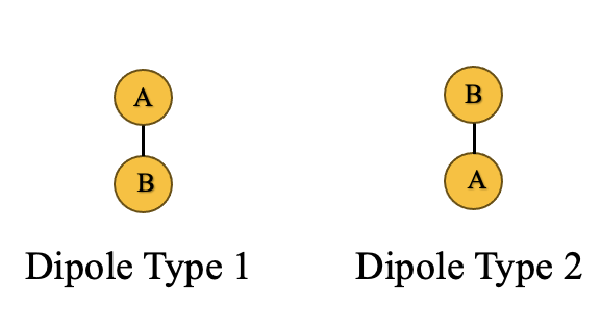}
    \caption[Possible dislocation dipole types]{The above figure depicts the possible \textbf{dislocation dipole types} for a dislocation dipole.}
    \label{dis-dipole-type}
\end{figure}

To fully characterize the spatial configuration of each dipole, we introduce the notion of \textbf{dipole type}. A \textbf{dipole type 1} refers to a configuration where the type A core lies above the type B core along the \( y \)-axis. Conversely, a \textbf{dipole type 2} configuration has the type B core positioned above the type A core (see Fig.~\ref{dis-dipole-type}).

To develop an effective one-dimensional model for tuning strain profiles and investigating flat band formation, we will first construct a statistical model in which the dislocation array approximates a honeycomb lattice. This approach is motivated by the fact that uniaxial periodic strain on a honeycomb lattice, such as graphene, can induce flat bands \cite{andrade2023}. By modeling the dislocation array in this way, we can then map the system onto an effective 1D model and analyze whether key features of the dipole distribution influence the emergence of flat electronic bands along specific directions, particularly the single direction present in our 1D approximation.

\section{Methodology}
We begin, as in our previous work \cite{Fall2025_FlatBands}, by considering the Fourier transform of the full dislocation array potential, given a set of dislocation core positions \( \{\mathbf{r}_{j}\} \), dipole vectors \( \{\mathbf{a}_{j}\} \), and dipole types \( t_j \in \{-1, +1\} \), where \( t_j \) distinguishes between \textbf{dipole type 1} and \textbf{dipole type 2} and \(A_{\mathbf{k}}\) is the dislocation potential in reciprocal space. The Fourier-transformed potential is given by
\begin{equation}
    \mathcal{F}\left\{A_{\text{FULL}}(\mathbf{r}_1,\ldots,\mathbf{r}_{N_{\text{dis}}};\, \mathbf{a}_1,\ldots,\mathbf{a}_{N_{\text{dis}}};\, t_1,\ldots,t_{N_{\text{dis}}})\right\}
    = A_{\mathbf{k}} \sum_{j = 1}^{N_{\text{dis}}} e^{-i \mathbf{k} \cdot \mathbf{r}_{j}} \left( 1 - e^{i t_j \mathbf{k} \cdot \mathbf{a}_{j}} \right),
\end{equation}
where \( N_{\text{dis}} \) is the number of dislocations, and \( \mathbf{r}_j \) denotes the position of the \textbf{type A} dislocation core in the \( j \)-th dipole. We assume a fixed dipole separation vector, so that \( \mathbf{a}_j = \mathbf{a} \) for all \( j \).

Our goal is to study the averaged second-order electron propagator for dislocation configurations sampled from a probability distribution over \( \{ \mathbf{r}_j \} \) and \( \{ t_j \} \), subject to the following conditions:
\newline
\newline
For each dislocation dipole located at \( \mathbf{r}_i = (x_i, y_i) \), where \( \mathbf{r}_i \) always denotes the position of the type A dislocation core within the dipole, and its \textbf{six} nearest neighbors \( \mathbf{r}_j = (x_j, y_j) \):
\begin{enumerate}
\item \label{constraint:neighbor_opposite}
If \( | y_i - y_j | < d \), then \( t_i = -t_j \). Any configuration that fails to satisfy this condition is assigned zero probability. Moreover, if a dipole has multiple neighbors that would impose conflicting requirements on \( t_i \), then the dipole cannot occupy that position.

\item \label{constraint:neighbor_opposite2}
If \( \big| y_i - y_j + (t_i - t_j) d \big| < d \), then \( t_i = -t_j \). As with the previous condition, any violation of this constraint leads to zero probability, and any conflicting constraints on \( t_i \) from different neighbors invalidate the configuration.

\item \label{constraint:uncorrelated}
If \( | y_i - y_j | > d \), then \( t_i \) and \( t_j \) are independent and equally likely to take values \( +1 \) or \( -1 \).

\end{enumerate}

where \( d \equiv |\mathbf{R}_{\mathrm{dip}}| \). These constraints effectively introduce a repulsive interaction between dislocation cores of the same type, ensuring that such cores are separated by at least a distance \( d \) along the \( y \)-axis. This promotes the tendency for neighboring dipoles to have opposite types (i.e., dipole type 1 and type 2), favoring the formation of dislocation quadrupoles when dipoles are near each other, a configuration that aligns with the expected physical arrangement of a pair of dislocation dipoles in order to minimize their total energy. Additionally, because the \( x \)-coordinates of neighboring dipoles are uncorrelated, it is improbable for them to share the same \( x \)-position. This statistical independence introduces effective lateral disorder along the \( x \)-direction.

\begin{figure}
    \centering
    \includegraphics[width=0.5\textwidth]{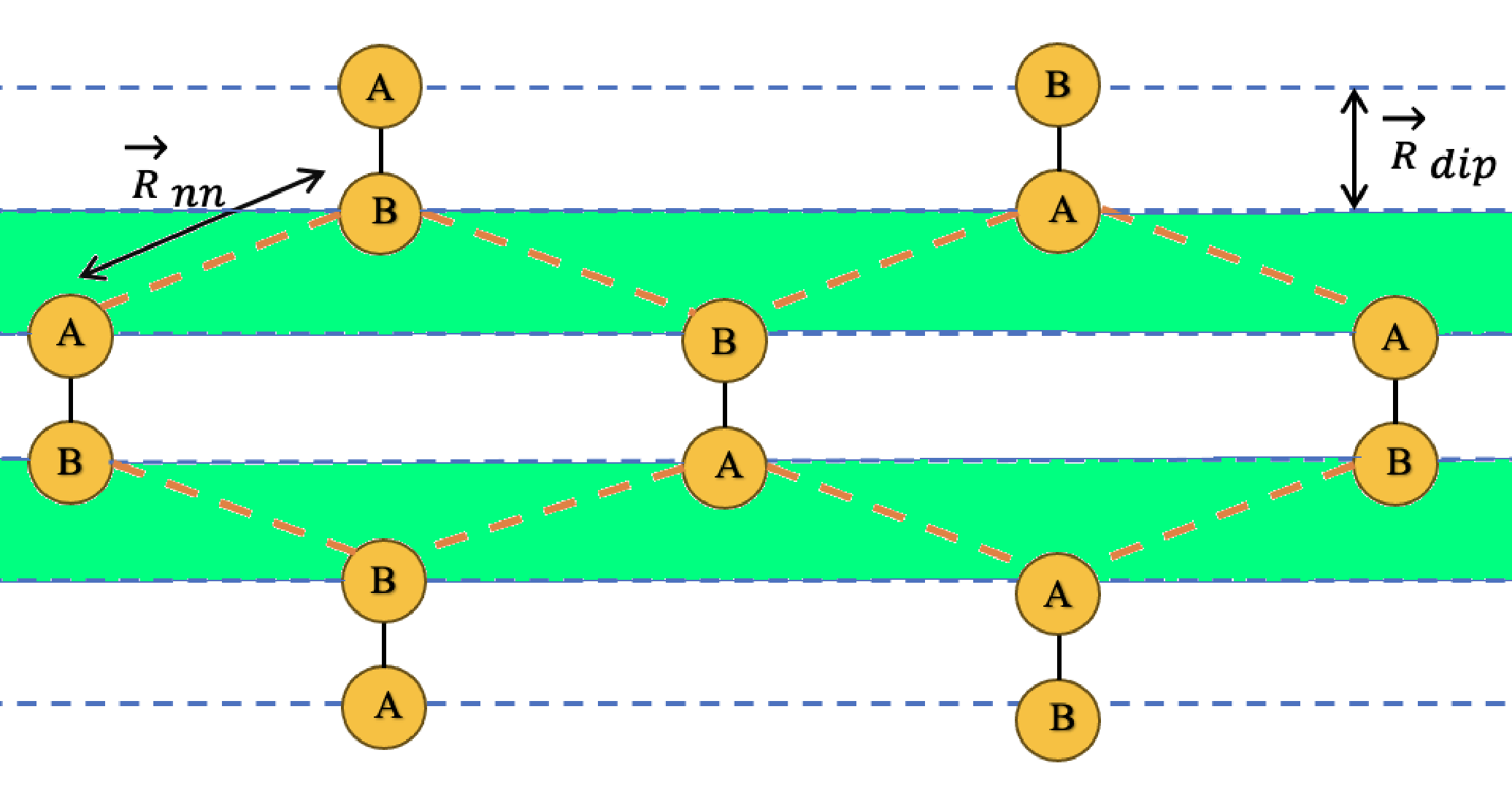}
    \caption[Dislocation dipole distribution with honeycomb-like pattern]{Example of a dislocation dipole distribution consistent with the probability conditions in \ref{constraint:neighbor_opposite}, \ref{constraint:neighbor_opposite2}, and \ref{constraint:uncorrelated}. The overall arrangement resembles a honeycomb-like structure. The shaded green region highlights an area where dipoles are effectively repelled from one another due to the imposed statistical constraints.}
    \label{honeycomb-dislocation}
\end{figure}

Together, these constraints generate a statistical ensemble in which the dislocation dipoles approximately form a honeycomb-like pattern, as illustrated in Fig.~\ref{honeycomb-dislocation}. Given these conditions, the probability distribution over the sets \( \{ \mathbf{r}_j \} \) and \( \{ t_j \} \) can be expressed as

\begin{equation}
    P(\{\mathbf{r}_j\}, \{t_j\}) \propto \exp\left\{ 
        -\lambda \bigg[ \sum_{\substack{ \langle ij \rangle \\ |y_i - y_j| < d } } 
        \left( 1 + t_i t_j \right) \ \ + \sum_{\substack{ \langle ij \rangle \\ |y_i - y_j + (t_{i} - t_{j})d \ | < d } } 
        \left( 1 + t_i t_j \right) \bigg] 
    \right\},
\end{equation}

where \(\langle ij \rangle\) denotes summation over each of the \textbf{six} nearest neighbors \( \{ \mathbf{r}_j \} \) of \(\mathbf{r}_i\), and \(\lambda\) is a parameter tending to infinity to enforce the constraints described above.

This form penalizes configurations where neighboring dipole cores of the same core type, separated vertically by less than \( d \), share the same dipole type (i.e., \( t_i t_j = +1 \)). As \( \lambda \to \infty \), such configurations become increasingly unlikely, effectively enforcing that neighboring dipoles have opposite types (\( t_i t_j = -1 \)). Consequently, the probability distribution becomes sharply concentrated around configurations consistent with the prescribed dislocation distribution.

To compute an electronic dispersion relation, translational invariance is required along at least one spatial direction(i.e. either the \( x \)-axis or the \( y \)-axis). Ideally, we would choose the \( x \)-axis, as the probability distribution is most uniform in that direction. However, the need to account for nearest-neighbor dipole configurations in our original probability model breaks this uniformity along \( x \), thereby obstructing translational invariance.

\begin{figure}[H]
    \centering
    \includegraphics[width=0.5\textwidth]{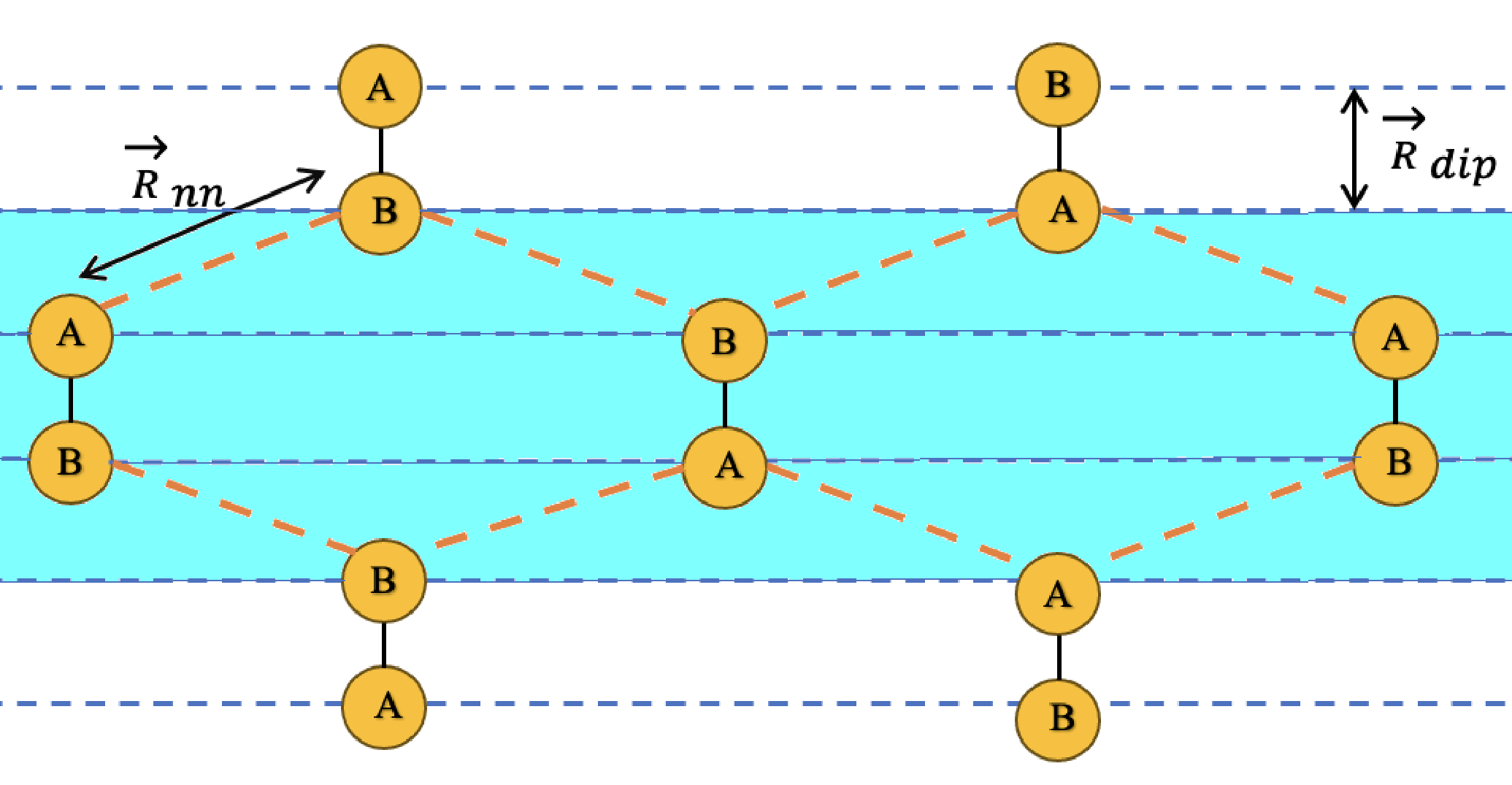}
    \caption[Constraint region for dislocation dipoles]{The blue shaded region illustrates the domain in which the reference dipole (centered in the figure) and all dipoles with \( y \)-coordinates within the interval \( [y_i - d,\, y_i + d] \) must satisfy the constraint that their dipole types sum to zero.}
    \label{constraint-update}
\end{figure}

To address this, we tweak the probability constraints to restore translational invariance along the \( x \)-axis, while still retaining some of its original features. This allows us to effectively reduce the system to a one-dimensional problem. Specifically, we impose the following condition:
\begin{equation}
\label{sum-constraint}
    \sum_{i} \sum_{j: y_j \in [y_i - d,\, y_i + d]} t_j = 0,
\end{equation}
 That is, for each dipole located at \( y_i \), we consider all dipoles with \( y \)-coordinates within the interval \( [y_i - d,\, y_i + d] \), and require that the sum of their dipole types \( t_j \), including the reference dipole at \( y_i \), is zero (see Fig.~\ref{constraint-update}).

In this formulation, the magnitude of the Burgers vector associated with each dipole determines the average amplitude of the induced strain field, while the value of \( d \) effectively controls the average wavelength of the strain modulation along the \( x \)-direction. By changing \( d \), one can generate different wavelength modulations, providing a handle for tuning the spatial profile of the strain landscape.

We can write the probability distribution over the dislocation positions \( \{ \mathbf{r}_{i} \} \) and dipole types \( \{ t_{i} \} \) to satisfy the imposed constraint in Eq.~\ref{sum-constraint} as  
\begin{equation}
 \label{approx-prob}
    P(\{ \mathbf{r}_{i} \}, \{ t_{i} \}) \propto \frac{1}{Z} \exp\bigg\{ - \lambda \sum_{i}^{N_{\mathrm{dis}}} \bigg[ \sum_{j}^{N_{\mathrm{dis}}} t_{j} \, K_{d}(y_{i},y_{j}) \bigg]^{2} \bigg\},
\end{equation}
where \( \lambda \) is a parameter tending to infinity, \( Z \) is the normalization factor, and \( K_{d} \) is the kernel or vertical window over which the sum of dipole types \( t_{j} \) is constrained to sum to zero.

 Following the approach of our previous work \cite{Fall2025_FlatBands}, we compute the effective one-dimensional dislocation–electron potential from the averaged second-order electron propagator. We begin with the expression for the averaged second-order propagator\cite{Fall2025_FlatBands, rammer2004} :
\begin{equation}
\begin{aligned}
    \left\langle G^{(2)}(\mathbf{p}, \mathbf{p}', E) \right\rangle &= \frac{1}{L^2} G_{0}(\mathbf{p},E) \frac{1}{L^2} \sum_{\mathbf{p}''} A_{\mathbf{p} - \mathbf{p}''} G_{0}(\mathbf{p}'',E) A_{\mathbf{p}'' - \mathbf{p}'} G_{0}(\mathbf{p}',E) \\
     &\quad \times \left\langle \sum_{m, n = 1}^{N_{\mathrm{dis}}} \left( 1 - e^{it_{m}(\mathbf{p - p''}) \cdot \mathbf{a}} \right) \left( 1 - e^{it_{n}(\mathbf{p'' - p'}) \cdot \mathbf{a} }  \right) e^{-i(\mathbf{p - p''}) \cdot \mathbf{r}_{m} } e^{-i(\mathbf{p'' - p'}) \cdot \mathbf{r}_{n}}  \right\rangle .
\end{aligned}
\end{equation}
Here, \( G_0(\mathbf{p}'', E) \) is the free electron propagator, dependent on both components of \( \mathbf{p}'' \). To reduce the problem to one dimension, we set the \( y \)-components of the incoming and outgoing momenta \( \mathbf{p} \) and \( \mathbf{p}' \) to zero, and sum over the intermediate transverse momentum \( p_y'' \), effectively integrating out the transverse degrees of freedom.

We further consider the case where \( x_{m} = x_{n} \equiv x \), while \( y_{m} \) and \( y_{n} \) are not constrained to have the same value. This yields an effective potential that depends only on the longitudinal momenta:
\begin{equation}
\begin{aligned}
\label{eff-pot-avg-prop}
\frac{1}{L^2} \sum_{m, n = 1}^{N_{\mathrm{dis}}} \frac{1}{L} \sum_{p_y''} &A_{\mathbf{p} - \mathbf{p}''} A_{\mathbf{p}'' - \mathbf{p}'} 
\left\langle \left( e^{i t_m (\mathbf{p} - \mathbf{p}'') \cdot \mathbf{a}} - 1 \right)
\left( e^{i t_n (\mathbf{p}'' - \mathbf{p}') \cdot \mathbf{a}} - 1 \right) e^{-ip_{y}''(y_{m} - y_{n}) }  \right\rangle \\ 
&\times \left\langle e^{-i(p_x - p_x')x } \right\rangle 
G_0(\mathbf{p}'', E)   \ \to\  \eta_{\mathrm{dis}} \, \left|A'_{p_x - p_x''}\right|^2 \, G_0(p_x'', E),
\end{aligned}
\end{equation}
where \( \eta_{\mathrm{dis}} \) is the dislocation density, defined as \( \eta_{\mathrm{dis}} \equiv \frac{N_{\mathrm{dis}}}{L^2} \). This factor is obtained by assuming \( N_{\mathrm{dis}} \approx (N_{\mathrm{dis}}^{y})^{2} \), i.e., the total number of dislocations is approximately equal to the square of the number of dislocations (or average number) along the \( y \)-axis. The right-hand side of Eq.~\eqref{eff-pot-avg-prop} no longer depends on \( p_y'' \), and \( A'_{p_x - p_x''} \) denotes the effective potential.

The angular brackets denote an average over dislocation positions \( \mathbf{r}_i \) and dipole types \( t_i \) using the probability distribution in Eq.~\ref{approx-prob}. This procedure produces a one-dimensional effective potential for the dislocation distribution.

\subsection{Computing the Effective Potential}
\label{1D-Potential}
To compute the effective one-dimensional potential, we first evaluate the average
\begin{equation}
\label{target-comp}
\left\langle \left( e^{i t_m (\mathbf{p} - \mathbf{p}'') \cdot \mathbf{a}} - 1 \right)
\left( e^{i t_n (\mathbf{p}'' - \mathbf{p}') \cdot \mathbf{a}} - 1 \right) e^{-ip_{y}''(y_{m} - y_{n}) }  \right\rangle  \left\langle e^{-i(p_x - p_x')x } \right\rangle .
\end{equation}
This requires knowledge of the marginal probability distribution of \(y_{1}, y_{2}, t_1, t_2\), and \(x\). The marginal distribution over \(x\) is straightforward: since dislocations are uniformly distributed along the \(x\)-axis, it is simply \(P(x) = 1/L\).

The marginal distribution over \(y_{1}, y_{2}, t_1, t_2\) is more challenging to compute directly from Eq.~\ref{approx-prob}. To proceed analytically, we adopt the ansatz
\begin{equation}
    P(y_1,y_2,t_1,t_2) = \frac{1}{Z} \exp\left[ -\lambda (t_1 + t_2)^2 K_d(y_1,y_2) \right] 
    = \frac{1}{Z} \exp\left[ -\lambda \left( 2 + 2t_1t_2 \right) \Theta(d - |y_2 - y_1|) \right],
\end{equation}
where the kernel \(K_d\) is taken to be the step function \(\Theta(d - |y_2 - y_1|)\). An alternative derivation of this marginal distribution obtained from Eq.~\ref{approx-prob} is given in Appendix. \ref{appendix-marginal}.

From this form, the normalization factor becomes
\begin{equation}
\begin{aligned}
    Z &= \sum_{t_1, t_2} \int_{0}^{L}dy_1 \int_{0}^{L} dy_{2} \ \exp\left[ -\lambda(2 + 2t_1t_2)\Theta(d - |y_2 - y_1|) \right] \\
      &= 4L^2 + (e^{-4\lambda} - 1)\left(4dL - 2d^2\right).
\end{aligned}    
\end{equation}
We now scale \(d\) with the system size by setting \(d = \beta L\), where \(0<\beta<1\). With this scaling, \(Z\) can be written as
\begin{equation}
    Z = 4L^2 Z_{\beta},
\end{equation}
where
\begin{equation}
    Z_{\beta} = 1 + \frac{\beta^2}{2} - \beta.
\end{equation}

We can now explicitly write out the average quantity we want to compute in Eq. \ref{target-comp} as 
\begin{equation}
\left\langle e^{-i(p_x - p_x')x} \right\rangle  
\sum_{t_1, t_2} \frac{G(t_1,t_2) }{4L^2Z_{\beta}} \int_{0}^{L} dy_1 \int_{0}^{L} dy_2 \ \exp\left[ -\lambda\left( 2 + 2t_1t_2\right) \Theta(d - |y_2 - y_1|) \right] e^{-ip_y''(y_2 - y_1)},
\end{equation}
where
\begin{equation}
    G(t_1, t_2) = \left( e^{i t_1 (\mathbf{p} - \mathbf{p}'') \cdot \mathbf{a}} - 1 \right)
\left( e^{i t_2 (\mathbf{p}'' - \mathbf{p}') \cdot \mathbf{a}} - 1 \right).
\end{equation}

Changing variables to \(w = y_1\) and \(r = y_2 - y_1\), and defining \(\alpha = \lambda(2 + 2t_1t_2)\), the expression becomes
\begin{equation}
\label{avg-b4-limit}
\left\langle e^{-i(p_x - p_x')x} \right\rangle  
\sum_{t_1, t_2} \frac{G(t_1,t_2) }{Z_{\beta}} \left[ \frac{1}{2L^2}\int_{0}^{L} dr \ (L - r)\cos(p_y''r) + \frac{1}{2L^2}\int_{0}^{L} dr \ (e^{-\alpha} - 1)(L - r)\cos(p_y''r) \Theta(d - r) \right].
\end{equation}

We now take the macroscopic limit \(N_{\mathrm{dis}} \to \infty\) and \(L \to \infty\). For the first term,
\begin{equation}
    \lim_{L \to \infty } \frac{1}{2L^2} \int_{0}^{L} dr \ (L - r)\cos(p_y''r) = \frac{1}{4}\delta(p_y'').
\end{equation}
For the second term, to obtain a well-defined limit we approximate the step function by a smooth Gaussian:
\begin{equation}
    \Theta(d - r) \approx \exp\left( - \frac{r^2}{2d^2} \right).
\end{equation}
This term becomes
\begin{equation}
    \frac{e^{-\alpha} - 1 }{2L^2} \int_{0}^{L} dr \ (L-r)\cos(p_y''r)\exp\left( - \frac{r^2}{2d^2} \right).
\end{equation}
With \(d = \beta L\) and \(r = Lu\), we have
\begin{equation}
    \lim_{L \to \infty} \frac{e^{-\alpha} - 1}{2} \int_{0}^{1} du \ (1 - u)\cos(p_y''Lu)e^{-u^2/(2\beta^2)}.
\end{equation}
For \(p_y'' \neq 0\), this is a highly oscillatory integral that goes to zero as \(L \to \infty \) except at \( p_y'' = 0 \). For \(p_y'' = 0\), it evaluates to
\begin{equation}
\begin{aligned}
    \frac{e^{-\alpha} - 1}{2} \left[ \left(e^{-1/(2\beta^2)} - 1\right)\beta^2 + \beta \sqrt{ \frac{\pi}{2} }\,\operatorname{erf}\left(\frac{1}{\beta \sqrt{2} } \right) \right].
\end{aligned}
\end{equation}

In the macroscopic limit, the original expression (i.e. Eq. \ref{avg-b4-limit}) reduces to
\begin{equation}
    \left\langle e^{-i(p_x - p_x')x} \right\rangle  
    \sum_{t_1, t_2} \frac{G(t_1,t_2) }{Z_{\beta}} \delta(p_y'') \left[ \frac{1}{4} +  \frac{e^{-\alpha} - 1}{4\pi} \left( \left(e^{-1/(2\beta^2)} - 1\right)\beta^2 + \beta \sqrt{ \frac{\pi}{2} }\,\operatorname{erf}\left(\frac{1}{\beta \sqrt{2} } \right) \right) \right].
\end{equation}

Finally, substituting into Eq.~\ref{eff-pot-avg-prop} and summing over \(t_1\) and \(t_2\) gives
\begin{equation}
    A_{\mathbf{k}} \to A_{k} \sqrt{\frac{1}{Z_{\beta}} \left[ \frac{3}{2} - \frac{D(\beta)}{\pi} + \left( \frac{D(\beta)}{\pi} - 2 \right)\cos(ka) + \frac{1}{2}\cos(2ka) \right] } \equiv A_k',
\end{equation}
where \(k\) is the crystal momentum, \(a\) is the dipole separation vector along the \(x\)-direction, and
\begin{equation}
    D(\beta) = \left(e^{-1/(2\beta^2)} - 1\right)\beta^2 + \beta \sqrt{ \frac{\pi}{2} }\,\operatorname{erf}\left(\frac{1}{\beta \sqrt{2} } \right).
\end{equation}

\subsection{Features of the Effective Potential}

To gain further insight into the effective potential computed above, we consider a mock system consisting of a random array of dislocation dipoles. All dislocation lines are taken to be aligned along the \(z\)-axis, with each dipole having a fixed separation in the \(x\)–\(y\) plane. In particular, we focus on the case where the slip-plane normal is  
\[
\mathbf{n} = \left( \frac{1}{\sqrt{2}}, \frac{1}{\sqrt{2}}, 0 \right),
\]
and the Burgers vector is  
\[
\mathbf{b} = \left( -\frac{b}{\sqrt{2}}, \frac{b}{\sqrt{2}}, 0 \right).
\]
For this configuration, the one-dimensional static dislocation potential along the \(x\)-axis takes the form \cite{Fall2025_FlatBands, li2018, li2019quantized, anderson2020theory, cai2016imperfections, stroh2021physics}
\begin{equation}
    A_k = f(\eta_{\mathrm{dis}}, Z, \nu, b) \left( \frac{1}{k^2 + k_{TF}^2} \right) \frac{1}{k},
\end{equation}
where the prefactor is given by
\begin{equation}
    f(\eta_{\mathrm{dis}}, Z, \nu, b) = i\, 2\pi Z \eta_{\mathrm{dis}} \frac{b(1 - 2\nu)}{2(\nu - 1)}.
\end{equation}
Here, \(b\) is the magnitude of the Burgers vector, \(\eta_{\mathrm{dis}}\) is the dislocation density, \(Z\) is the effective nuclear charge, \(\nu\) is the Poisson ratio, and \(k_{TF}\) is the Thomas–Fermi screening wavevector.

\begin{figure}[H]
    \centering
    \includegraphics[width=0.6\textwidth]{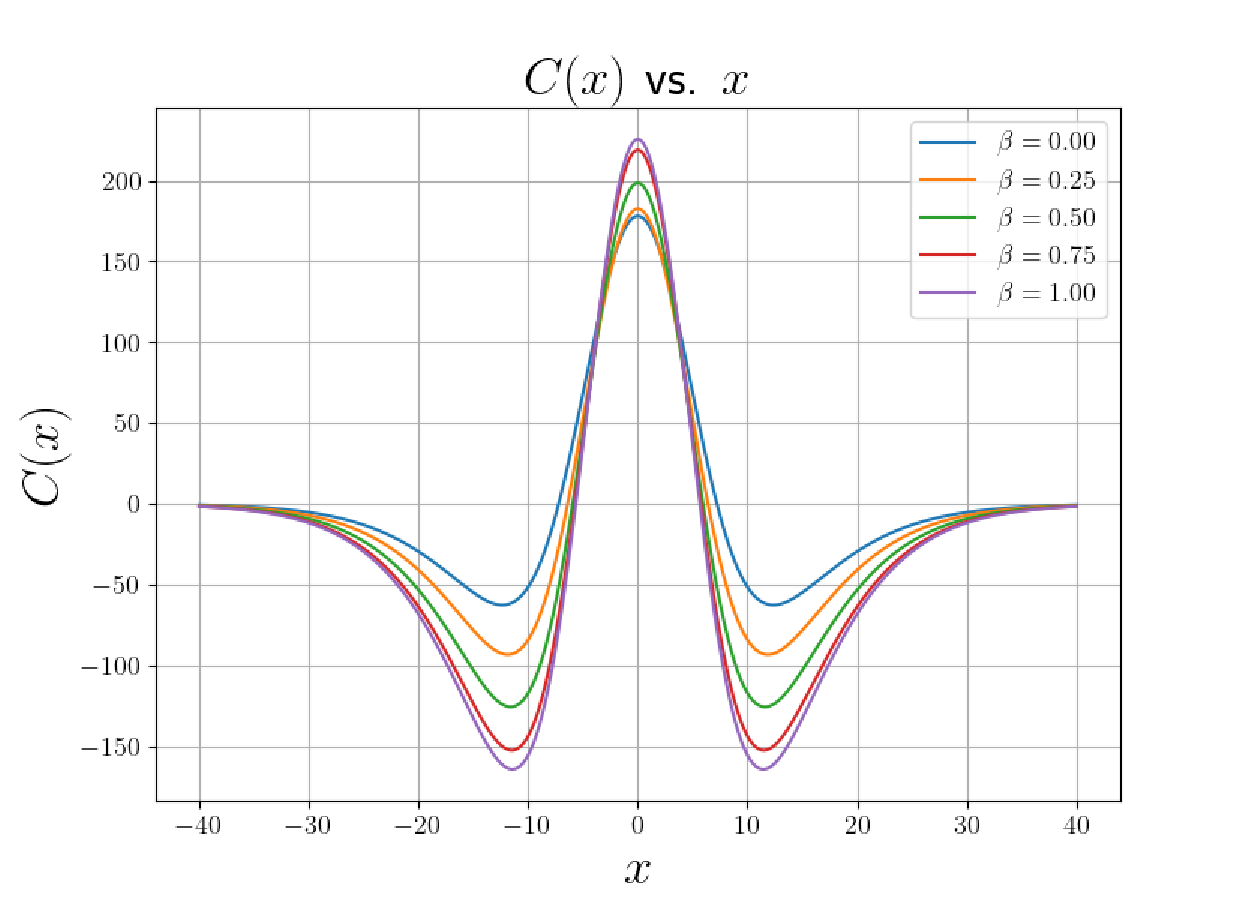}
    \caption[Autocorrelation of effective scattering potential]{Autocorrelations of the effective scattering potential \( A_k'\) in real space for different \(\beta \) values.}
    \label{fig:Auto_eff_pot}
\end{figure}

\begin{figure}[H]
    \centering
    \includegraphics[width=0.6\textwidth]{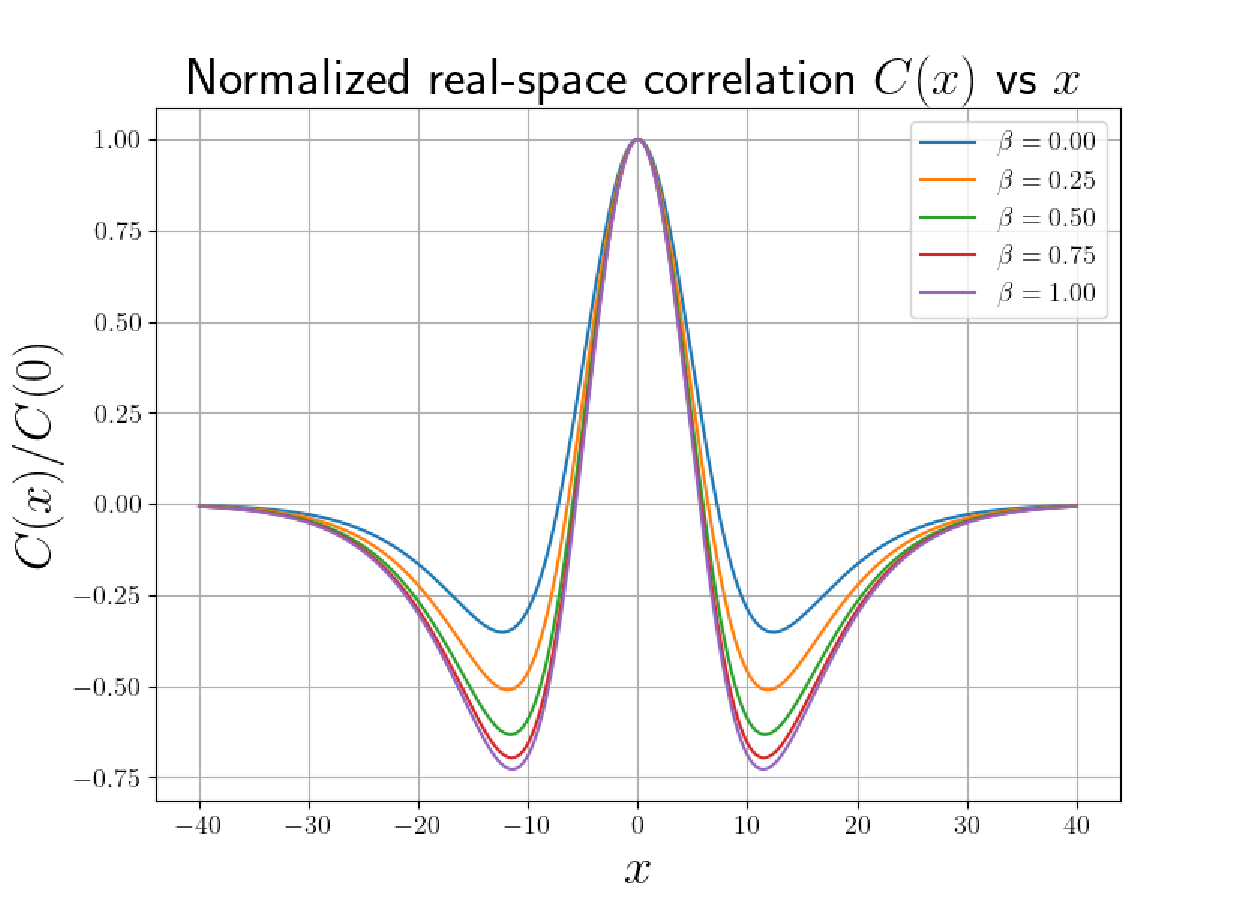}
    \caption[Normalized autocorrelation of effective potential]{Autocorrelations of the effective potential normalized by their peaks for different \(\beta\) values.}
    \label{fig:Auto_eff_pot_norm}
\end{figure}

To study how the effective potential \(A'_k\) changes with \(\beta\), we compute the autocorrelation of the effective potential,
\[
(A' \ast A')(x) \equiv C(x),
\]
where \(\ast\) denotes convolution. Figures~\ref{fig:Auto_eff_pot} and \ref{fig:Auto_eff_pot_norm} show the autocorrelation \(C(x)\) in real space for the parameters \(a = 7.2\), \(k_{TF} = 0.2336\), and with \(f(\eta_{\mathrm{dis}}, Z, \nu, b)\) set to unity, since it only acts as a constant multiplicative factor. Figure~\ref{fig:Auto_eff_pot} presents the raw autocorrelation curves for several values of \(\beta\), while Fig.~\ref{fig:Auto_eff_pot_norm} shows the same data normalized by their peak values to facilitate direct comparison of the width.

\begin{figure}[H]
    \centering
    \includegraphics[width=0.6\textwidth]{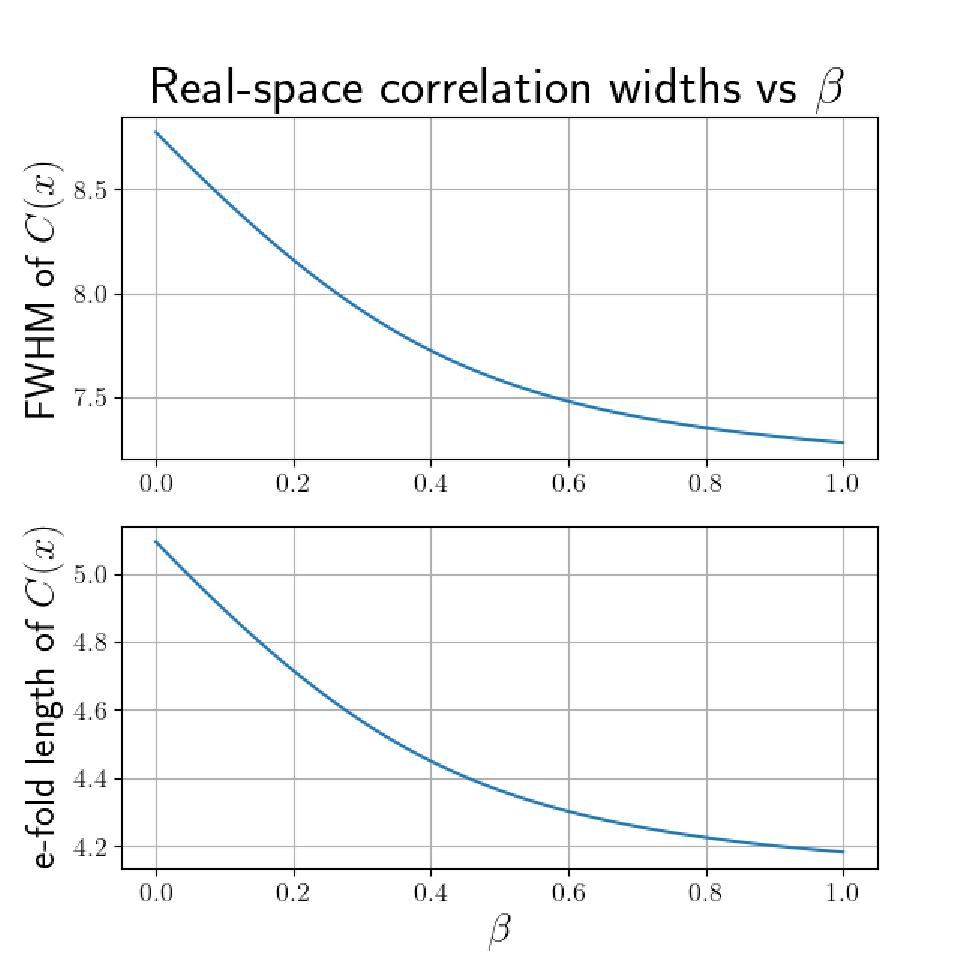}
    \caption[Autocorrelation widths vs.~\(\beta\)]{Widths of the autocorrelations as a function of \( \beta \) using full width at half maximum (FWHM) and e-fold decay metrics.}
    \label{fig:width_eff_pot}
\end{figure}

From these results, we observe that increasing \(\beta\) causes the peak of the autocorrelation to grow while its width decreases. This trend is quantified in Fig.~\ref{fig:width_eff_pot}, where the autocorrelation width is plotted as a function of \(\beta\) using both the full width at half maximum (FWHM) and the e-fold decay metric (width at which the function decreases to approximately \(37\%\) of its maximum). The data show that for large \(\beta\), the system exhibits greater variation in strain along the \(x\)-axis. A plausible explanation is that when \(\beta\) is large, dislocation dipoles are less likely to form quadrupoles when they are in close proximity, instead adopting more random configurations. This increased randomness in arrangement leads to stronger strain variations along the \(x\)-axis. As expected, \(d\) (or equivalently \(\beta\)) directly controls the average strain modulation in the longitudinal direction.

\section{Conclusion}
In order to understand the origin of the flat band formation discovered in our previous work \cite{Fall2025_FlatBands}, we reduced the dimensionality of the system by developing an effective one-dimensional dislocation potential. The effective potential was derived by averaging over different spatial distributions of dislocations along the singular direction. In addition, we introduced the parameter \(d = \beta L\), which scales linearly with the system, that can be adjusted to control the average strain modulation along this direction. Our effective model paves the way for a more comprehensive understanding of how dislocation strain can induce electronic flat bands in solid-state materials.

\section*{Acknowledgment}

We thank Ira Rothstein for his insightful comments; ARO (MURI W911NF-25-2-0164), NSF(2108784), and the GEM, McGaw, and Adamson Fellowships to Aziz Fall for financial support; and NSF ACCESS (PHY240235) for computing resources provided by the Pittsburgh Supercomputing Center.

\section*{Funding Data}

\begin{itemize}
    \item NSF(2108784)
    \item ARO (MURI W911NF-25-2-0164)
\end{itemize}

\section*{Conflict of Interest}

There are no conflicts of interest.

\section*{Data Availability Statement}

The data that support the findings of this article are openly available \cite{githubGitHubAzizfallDislocation_Induced_Flat_Bands}, and the computational results presented in this work can be reproduced using the provided codebase, also available at \cite{githubGitHubAzizfallDislocation_Induced_Flat_Bands}.

\appendix 
\section{Alternative Approach to Marginal Distribution}
\label{appendix-marginal}

In this section we demonstrate an alternative approach to calculating the marginal probability of the joint distribution \(t_{1}, t_{2}, y_{1}, y_{2}\) from Eq.~\ref{approx-prob}. We begin by rewriting the probability distribution from Eq.~\ref{approx-prob} as  
\begin{equation}
P( \{\mathbf{r}_{i}\}, \{t_{i}\}) = \frac{1}{Z} \exp \bigg\{ - \lambda \sum_{ijk}^{N} t_{i}t_{j}K_{d}(y_{k}, y_{i})K_{d}(y_{k}, y_{j})  \bigg\}.
\end{equation}

Next, we isolate the terms involving \(t_{1}, t_{2}, y_{1}, y_{2}\):  
\begin{equation}
\begin{aligned}
P(\{\mathbf{r}_{i}\}, \{t_{i}\}) 
  &= \frac{1}{Z} \exp \Bigg\{ -2\lambda \Bigg( 
     \sum_{\substack{k \\ k \ne 1 \\ k \ne 2}}^{N} 
     \Big[ t_{1}t_{2}K_d(y_k,y_1)K_d(y_k,y_2) \\
  &\qquad + t_1t_kK_d(y_2,y_1)K_d(y_2,y_k) 
     + t_2t_kK_d(y_1,y_k)K_d(y_1,y_2) \Big] \\
  &\qquad + (1 + 2t_1t_2)K_d(y_1,y_2) + O(t_3,...,t_N; y_3,...y_N) \Bigg) \Bigg\},
\end{aligned}
\end{equation}
where \(O(t_3,...,t_N; y_3,...y_N)\) collects the remaining terms not involving \(t_{1}, t_{2}, y_{1}, y_{2}\).  

We again take the kernel \(K_d\) to be the step function  
\[
K_d(y_i,y_j) = \Theta(d - |y_i - y_j|).
\]  
To proceed, we neglect the term \(O(t_3,...,t_N; y_3,...y_N)\). This yields an approximate marginal distribution of Eq.~\ref{approx-prob}. Integrating the numerator over \(y_k\), and computing the normalizing factor \(Z\) under the assumption \(d \ll L\), gives  
\begin{equation}
P(y_1, y_2, t_1, t_2) =  \frac{1}{Z} \exp \bigg\{ - \lambda[4t_1t_2 + 2] \Theta(d - |y_2 - y_1|)    \bigg\} \bigg[ 2L  + dI \bigg]^{N-2},
\end{equation}
with  
\begin{equation}
\begin{aligned}
    Z = \sum_{t_1, t_2} \bigg\{ (2L)^{N-2}(L^2 - 4Ld) + 2Le^{-\lambda[4t_1t_2 + 2]} \frac{(2L + H_1 + dG_1)^{N-1} - (2L + H_1)^{N-1}}{G_1(N - 1)}  \\ 
    + 2L \frac{(2L + H_2 + 2dG_2)^{N-1} - (2L + H_2 + dG_2)^{N-1}}{G_2(N - 1)}    \bigg\}.
\end{aligned}
\end{equation}

The parameters are defined as  
\begin{equation}
\begin{aligned}
        G_{1} &\equiv \bigg[ -2 + (e^{-2 \lambda t_{1}} + e^{2 \lambda t_{1}}) + (e^{-2 \lambda t_{2}} + e^{2 \lambda t_{2}}) \\
        &\qquad - (e^{-\lambda (2t_{2} + 2t_{1}t_{2} + 2t_{1})} + e^{\lambda (2t_{2} - 2t_{1}t_{2} + 2t_{1})}) \bigg],
\end{aligned}
\end{equation}

\begin{equation}
        H_{1} \equiv - 4 + 2\big( e^{-\lambda(2t_{2} + 2t_{1}t_{2} + 2t_{1})} + e^{\lambda(2t_{2} - 2t_{1}t_{2} + 2t_{1}) } \big),
\end{equation}

\begin{equation}
     G_{2} \equiv 2 - 2e^{-2\lambda t_{1}t_{2}},
\end{equation}

\begin{equation}
        H_{2} \equiv - 4 + 4e^{-2\lambda t_{1} t_{2}},
\end{equation}

and  
\begin{equation}
\begin{aligned}
I &= 2(\Delta - 4) + (2 - \Delta)\big(e^{-2\lambda t_{2}\Theta(d - |y_{2} - y_{1}|)} + e^{2 \lambda t_{2} \Theta(d - |y_{2} - y_{1}|)}\big) \\
&\quad + (2 - \Delta)\big(e^{-2\lambda t_{1}\Theta(d - |y_{2} - y_{1}|)} + e^{2 \lambda t_{1} \Theta(d - |y_{2} - y_{1}|)}\big) \\
&\quad + \Delta \Big[ e^{-\lambda(2t_{2}\Theta(d - |y_{2} - y_{1}|) + 2t_{1}t_{2} + 2t_{1}\Theta(d - |y_{2} - y_{1}|))} \\
&\qquad + e^{\lambda(2t_{2}\Theta(d - |y_{2} - y_{1}|) - 2t_{1}t_{2} + 2t_{1}\Theta(d - |y_{2} - y_{1}|)) }  \Big],
\end{aligned}
\end{equation}

with  
\begin{equation}
    \Delta = \max\left(0, 2 - \frac{|y_{2} - y_{1}|}{d}\right).
\end{equation}

In the limit \(d \ll L\), we approximate  
\begin{equation}
    [2L + dI]^{N-2} = (2L)^{N-2}\bigg[1 + \frac{dI}{2L}  \bigg]^{N-2} 
    \approx (2L)^{N-2} \exp \bigg[ I\frac{(N-2)d }{2L} \bigg].
\end{equation}

This motivates defining a dislocation density parameter
\begin{equation}
    w = \frac{Nd}{L}.
\end{equation}
The probability function will converge provided that \(d\) scales slower than or equal to \(1/L^{2}\).  

\medskip  
\noindent
\textbf{Key distinction:} In section~\ref{1D-Potential}, the strain profile of the effective potential was tuned by varying a single parameter \(d\), which controls the local interaction range between dislocations. In contrast, the present formulation tunes the strain profile through the collective density parameter \(w\), which incorporates the contributions of all dislocations in the system. This represents a different way to compute the effective potential by using an alternative marginal distribution.

\end{document}